\documentclass[12pt]{article}
\catcode`\@=11
\def\numberbysection{\@addtoreset{equation}{section}
\def\theequation{\thesection.\arabic{equation}}}
\numberbysection

\newcommand{\abs}[1]{\vert#1\vert}
\newcommand{\beq}{\begin{equation}}
\newcommand{\beqa}{\begin{eqnarray}}
\newcommand{\comport}[2]{\mathrel{\mathop{#1}\limits_{#2}^{}}}
\renewcommand{\d}{{\rm d}}
\newcommand{\dq}{\!{\frac{\d^D\q}{(2\pi)^D}}}
\newcommand{\dpar}{\partial}
\newcommand{\e}{{\rm e}}
\newcommand{\eeq}{\end{equation}}
\newcommand{\eeqa}{\end{eqnarray}}
\newcommand{\eq}{{\rm eq}}
\newcommand{\eps}{\varepsilon}
\newcommand{\eqcr}{{\rm eq,c}}
\newcommand{\fd}{fluc\-tu\-a\-tion-dis\-si\-pa\-tion }
\newcommand{\frad}[2]{\displaystyle{\displaystyle#1\over\displaystyle#2}}
\newcommand{\F}{{\rm F}}
\renewcommand{\H}{{\cal H}}
\renewcommand{\i}{{\rm i}}
\renewcommand{\L}{{\rm L}}
\newcommand{\mf}{{\rm MF}}
\newcommand{\mean}[1]{\langle#1\rangle}
\newcommand{\qea}{q_{\rm EA}}
\newcommand{\reg}{{\rm reg}}
\newcommand{\sg}{{\rm sg}}
\newcommand{\taueq}{\tau_{\rm eq}}
\newcommand{\TRM}{{\rm TRM}}
\newcommand{\q}{{\bf q}}
\newcommand{\0}{{\bf 0}}
\newcommand{\x}{{\bf x}}
\newcommand{\X}{{\cal X}}
\newcommand{\y}{{\bf y}}
\newcommand{\z}{Z}
\begin{document}
\centerline{\Large\bf Response of non-equilibrium systems at criticality:}
\vspace{.3cm}
\centerline{\Large\bf Ferromagnetic models in dimension two and above}
\vspace{1cm}
\centerline{\large
by C.~Godr\`eche$^{a,}$\footnote{godreche@spec.saclay.cea.fr}
and J.M.~Luck$^{b,}$\footnote{luck@spht.saclay.cea.fr}}
\vspace{1cm}
\centerline{$^a$Service de Physique de l'\'Etat Condens\'e,
CEA Saclay, 91191 Gif-sur-Yvette cedex, France}
\vspace{.1cm}
\centerline{$^b$Service de Physique Th\'eorique,
CEA Saclay, 91191 Gif-sur-Yvette cedex, France}
\vspace{1cm}
\begin{abstract}
We study the dynamics of ferromagnetic spin systems quenched
from infinite temperature to their critical point.
We show that these systems are aging in the long-time regime,
i.e., their two-time autocorrelation
and response functions and associated \fd ratio
are non-trivial scaling functions of both time variables.
This is exemplified by the exact analysis of the spherical model
in any dimension $D>2$,
and by numerical simulations on the two-dimensional Ising model.
We show in particular that, for $1\ll s$ (waiting time)
$\ll t$ (observation time),
the \fd ratio possesses a non-trivial limit value $X_\infty$,
which appears as a dimensionless amplitude ratio,
and is therefore a novel universal characteristic of non-equilibrium
critical dynamics.
For the spherical model, we obtain $X_\infty=1-2/D$ for $2<D<4$,
and $X_\infty=1/2$ for $D>4$ (mean-field regime).
For the two-dimensional Ising model we measure $X_\infty\approx0.26\pm0.01$.
\end{abstract}
\vfill
\noindent To be submitted for publication to Journal of Physics A
\hfill S/00/006
\vskip -6pt
\noindent P.A.C.S.: 02.50.Ey, 05.40.+j, 61.43.Fs, 75.50.Lk
\hfill T/00/004
\newpage

\section{Introduction}

Consider a ferromagnetic model, without quenched randomness,
evolving from a disordered initial state, according to some dynamics
at fixed temperature~$T$.
In the high-temperature paramagnetic phase $(T>T_c)$,
the system relaxes exponentially to equilibrium.
At equilibrium, two-time quantities such as the
autocorrelation function $C(t,s)$ or the response function $R(t,s)$
only depend on the time difference $\tau=t-s$,
where~$s$ (waiting time) is smaller than~$t$ (observation time),
and both quantities are simply related to each other by the \fd theorem
\beq
R_\eq(\tau)=-\frac{1}{T}\,\frac{\d C_\eq(\tau)}{\d\tau}.
\label{fdt}
\eeq
In the low-temperature phase $(T<T_c)$ the system undergoes phase ordering.
In this non-equilibrium situation, $C(t,s)$ and $R(t,s)$
are non-trivial functions of both time variables,
which only depend on their ratio at late times,
i.e., in the self-similar domain growth (or coarsening) regime~\cite{bray}.
This behavior is usually referred to as aging~\cite{revue}.
Moreover, no such simple relation as eq.~(\ref{fdt}) holds
between correlation and response, i.e.,
$R(t,s)$ and $\dpar C(t,s)/\dpar s$ are no longer proportional.
It is then natural to characterize the distance to equilibrium
of an aging system by the so-called \fd ratio~\cite{revue,ckp,ck}
\beq
X(t,s)=\frac{T\,R(t,s)}{\,\frad{\dpar C(t,s)}{\dpar s}\,}.
\label{dx}
\eeq
In recent years,
several works~\cite{revue,ckp,ck,cd,x1,barrat,berthier,autres,zkh}
have been devoted to the study of the \fd ratio
for systems exhibiting domain growth,
or for aging systems such as glasses and spin glasses,
showing that in the low-temperature phase
$X(t,s)$ turns out to be a non-trivial function of its two arguments.
In particular, for domain-growth systems,
analytical and numerical studies indicate that the limit \fd ratio,
\beq
X_\infty=\lim_{s\to\infty}\lim_{t\to\infty}X(t,s),
\label{xinfdef}
\eeq
vanishes throughout the low-temperature phase~\cite{x1,barrat,berthier}.

However, to date, only very little attention has been devoted
to the response function $R(t,s)$, and
\fd ratio $X(t,s)$, for non-equilibrium systems {\it at criticality}.
{}From now on, we will only have in mind ferromagnetic systems
without quenched randomness.
For instance one may wonder whether there exists,
for a given model, a well-defined limit $X_\infty$ at $T=T_c$,
different from its trivial value $X_\infty=0$ in the low-temperature phase,
and to what extent $X_\infty$ is universal.
Indeed a priori, for a system such as a ferromagnet,
quenched from infinitely high temperature to its critical point,
the limit \fd ratio $X_\infty$ at $T=T_c$
(if it exists) may take any value between $X_\infty=1$ ($T>T_c$: equilibrium)
and $X_\infty=0$ ($T<T_c$: domain growth).

The only cases of critical systems for which the \fd ratio has been
considered are, to our knowledge, the models of ref.~\cite{ckp}
(random walk, free Gaussian field,
and two-dimensional X-Y model at zero temperature)
which share the limit \fd ratio $X_\infty=1/2$,
and the backgammon model, a mean-field model for which $T_c=0$,
where it has been shown that $X_\infty=1$,
up to a large logarithmic correction,
for both energy fluctuations and density fluctuations~\cite{fr97,gl99}.

In a recent companion paper~\cite{gl}, we have determined the \fd ratio
$X(t,s)$ for the Glauber-Ising chain, another model for which $T_c=0$.
In particular, the limit \fd ratio was found to be $X_\infty=1/2$
(see also ref.~\cite{zan}).
In the present work we investigate the non-equilibrium correlation
and response functions and the associated \fd ratio
in generic ferromagnetic models at their critical point.
We first present (in section~2) an analytical study of the spherical model
in arbitrary dimension.
We then turn (in section~3) to a scaling analysis of the generic case,
and to numerical simulations on the two-dimensional Ising model.
One salient outcome of these joint works is the realization
that the limit \fd ratio $X_\infty$
is a novel universal characteristic of critical dynamics,
intrinsically related to non-equilibrium initial situations.

The present paper is written in a self-contained fashion.
For the spherical model,
though our main intention lies in the study of non-equilibrium
dynamics at the critical point,
we shall present the three situations $T>T_c$, $T=T_c$, and $T<T_c$
in parallel.
The latter case has already been the subject of a number of investigations,
for both correlation and response~\cite{cd,zkh,bray}.
Results on the scaling behavior of the two-time autocorrelation
function at $T_c$ can be found in ref.~\cite{janssen}.
For the two-dimensional Ising model,
several numerical works have already been devoted
to its non-equilibrium dynamics in the low-temperature phase,
concerning both correlations~\cite{bray,fisher} and response~\cite{barrat}.
We will therefore restrict our numerical study to the dynamics at the
critical point.

\section{The spherical model}

\subsection{Langevin dynamics}

The ferromagnetic spherical model was introduced by
Berlin and Kac~\cite{berlin}, as an attempt to simplify the Ising model.
It is solvable in any dimension, yet possesses non-trivial critical
properties~\cite{berlin,baxter}.
Consider a lattice of points of arbitrary dimension~$D$,
chosen to be hypercubic for simplicity, with unit lattice spacing.
The spins $S_\x$, situated at the lattice vertices $\x$,
are real variables subject to the constraint
\beq
\sum_\x S_\x^2=N,
\label{constr}
\eeq
where~$N$ is the number of spins in the system.
The Hamiltonian of the model reads
\beq
\H=-\sum_{(\x,\y)}S_\x S_\y,
\label{ham}
\eeq
where the sum runs over pairs of neighboring sites.

Throughout the following, we assume that the system is homogeneous,
i.e., invariant under spatial translations.
This holds for a finite sample with periodic boundary conditions,
and (at least formally) for the infinite lattice.
We also assume that the initial state of the system
at $t=0$ is the infinite-temperature equilibrium state.
This state is fully disordered, in the sense that spins are uncorrelated.
The dynamics of the system is given by
the stochastic differential Langevin equation
\beq
\frac{\d S_\x}{\d t}=\sum_{\y(\x)}S_\y-\lambda(t)S_\x+\eta_\x(t).
\label{lan}
\eeq
The first term, where $\y(\x)$ denotes the $2D$ first neighbors
of the site $\x$, is equal to the gradient $-\dpar\H/\dpar S_\x$,
while $\lambda(t)$ is a Lagrange multiplier
ensuring the constraint~(\ref{constr}), which we choose to parameterize as
\beq
\lambda(t)=2D+z(t),
\label{zdef}
\eeq
and $\eta_\x(t)$ is a Gaussian white noise with correlation
\beq
\mean{\eta_\x(t)\eta_\y(t')}=2T\,\delta_{\x,\y}\,\delta(t-t').
\eeq

Equation~(\ref{lan}) can be solved in Fourier space.
Defining the spatial Fourier transform by the formulas
\beq
f^\F(\q)=\sum_\x f_\x\,\e^{-\i\q.\x},\qquad
f_\x=\int\dq\,f^\F(\q)\,\e^{\i\q.\x},
\eeq
where
\beq
\int\dq=\int_{-\pi}^{\pi}\frac{\d q_1}{2\pi}
\cdots\int_{-\pi}^{\pi}\frac{\d q_D}{2\pi}
\eeq
is the normalized integral over the first Brillouin zone, we obtain
\beq
\frac{\dpar S^\F(\q,t)}{\dpar t}=-[\omega(\q)+z(t)]S^\F(\q,t)+\eta^\F(\q,t),
\label{lanq}
\eeq
where
\beq
\omega(\q)=2\sum_{a=1}^D(1-\cos q_a)\comport{\approx}{\q\to\0}\q^2,
\label{omdef}
\eeq
and
\beq
\mean{\eta^\F(\q,t)\eta^\F(\q',t')}
=2T\,(2\pi)^D\,\delta^D(\q+\q')\,\delta(t-t').
\label{etavar}
\eeq
The solution to eq.~(\ref{lanq}) reads
\beq
S^\F(\q,t)=\e^{-\omega(\q)t-\z(t)}
\left(S^\F(\q,t=0)+\int_0^t\e^{\omega(\q)t_1+\z(t_1)}\eta^\F(\q,t_1)
\,\d t_1\right),
\label{sq}
\eeq
with
\beq
\z(t)=\int_0^t z(t_1)\,\d t_1.
\eeq

\subsection{Equal-time correlation function}

Our first goal is to compute the equal-time correlation function
\beq
C_{\x-\y}(t)=\mean{S_\x(t)S_\y(t)},
\label{crt}
\eeq
which is a function of the separation $\x-\y$, by translational invariance.
We have in particular
\beq
C_\0(t)=\mean{S_\x(t)^2}=1,
\label{c0t}
\eeq
because of the spherical constraint~(\ref{constr}), and
\beq
C_\x(t=0)=\delta_{\x,\0},
\label{cinit}
\eeq
reflecting the absence of correlations in the initial state.
In eq.~(\ref{crt}), the brackets denote the average
over the ensemble of infinite-temperature initial configurations
and over the thermal histories (realizations of the noise).

In Fourier space the equal-time correlation function is defined by
\beq
\mean{S^\F(\q,t)S^\F(\q',t)}=(2\pi)^D\,\delta^D(\q+\q')\,C^\F(\q,t).
\label{cqdef}
\eeq
Using the expression~(\ref{sq}),
averaging it over the white noise $\eta^\F(\q,t)$
with variance given by eq.~(\ref{etavar}), and imposing the condition
\beq
C^\F(\q,t=0)=1
\eeq
implied by eq.~(\ref{cinit}), we obtain
\beq
C^\F(\q,t)=\e^{-2\omega(\q)t-2\z(t)}
\left(1+2T\int_0^t\e^{2\omega(\q)t_1+2\z(t_1)}\,\d t_1\right).
\label{cq}
\eeq

At this point, we are naturally led to introduce two functions, $f(t)$ and
$g(T,t)$, which play a central role in the following developments.

The function $f(t)$ is explicitly given by
\beq
f(t)=\int\dq\,\e^{-2\omega(\q)t}=\left(\e^{-4t}I_0(4t)\right)^D
\comport{\approx}{t\to\infty}(8\pi t)^{-D/2},
\label{fdef}
\eeq
where
\beq
I_0(z)=\int\frac{\d q}{2\pi}\,\e^{z\cos q}
\comport{\approx}{z\to\infty}\,(2\pi z)^{-1/2}\,\e^z
\eeq
is the modified Bessel function.

The function
\beq
g(T,t)=\e^{2\z(t)}
\label{gdef}
\eeq
is related to $f(t)$ by the constraint~(\ref{c0t}), namely
\beq
\int\dq\,C^\F(\q,t)
=\frac{1}{g(T,t)}\left(f(t)+2T\int_0^tf(t-t_1)g(T,t_1)\,\d t_1\right)=1,
\label{constrq}
\eeq
which yields a linear Volterra integral equation for $g(T,t)$~\cite{cd}, namely
\beq
g(T,t)=f(t)+2T\int_0^tf(t-t_1)g(T,t_1)\,\d t_1.
\eeq
This equation can be solved using temporal Laplace transforms, denoted by
\beq
f^\L(p)=\int_0^\infty f(t)\,\e^{-pt}\,\d t.
\eeq
We obtain
\beq
g^\L(T,p)=\frac{f^\L(p)}{1-2Tf^\L(p)},
\label{fglap}
\eeq
with
\beq
f^\L(p)=\int\dq\,\frac{1}{p+2\omega(\q)}.
\label{fldef}
\eeq
The dependence of $g^\L(T,p)$ on temperature appears explicitly in
eq.~(\ref{fglap}).

We now present an analysis of the long-time behavior of the function $g(T,t)$,
considering successively the paramagnetic phase ($T>T_c$),
the ferromagnetic phase ($T<T_c$), and the critical point ($T=T_c$).
To do so, we shall extensively utilize eq.~(\ref{fglap}).
We therefore investigate first the function $f^\L(p)$,
as given in eq.~(\ref{fldef}).
This function has no closed-form expression, except in one and two dimensions:
\beqa
D=1:&&f^\L(p)=\frac{1}{\sqrt{p(p+8)}},\cr
D=2:&&f^\L(p)=\frac{2}{\pi\abs{p+8}}
\,{\bf K}\!\left(\frac{8}{\abs{p+8}}\right),
\eeqa
where ${\bf K}$ is the complete elliptic integral.
Together with the definition~(\ref{omdef}),
eq.~(\ref{fldef}) implies that $f^\L(p)$ is analytic in the complex $p$-plane
cut along the real interval $[-8D,0]$.
The behavior of $f^\L(p)$ in the vicinity of the branch point at $p=0$
can be analyzed heuristically as follows.
The asymptotic behavior of $f(t)$ given in eq.~(\ref{fdef})
suggests that its Laplace transform has a universal singular part:
\beq
f^\L_\sg(p)\comport{\approx}{p\to0}(8\pi)^{-D/2}\Gamma(1-D/2)p^{D/2-1},
\label{fsg}
\eeq
while there also exists a regular part of the form
\beq
f^\L_\reg(p)=A_1-A_2p+A_3p^2+\cdots,
\label{freg}
\eeq
where
\beq
A_k=\int\dq\,\frac{1}{(2\omega(\q))^k}
\eeq
are non-universal (lattice-dependent) numbers,
given in terms of integrals which are convergent for $D-2k>0$.
For instance $A_1$ only exists for $D>2$, and so on.
Equations~(\ref{fsg}) and~(\ref{freg}) jointly determine
the small-$p$ behavior of $f^\L(p)$, as a function of the dimensionality~$D$:
\beqa
D<2:&&f^\L(p)\approx(8\pi)^{-D/2}\Gamma(1-D/2)p^{-(1-D/2)},\cr
2<D<4:&&f^\L(p)\approx A_1-(8\pi)^{-D/2}\abs{\Gamma(1-D/2)}p^{D/2-1},\cr
D>4:&&f^\L(p)\approx A_1-A_2p.
\label{flist}
\eeqa
These expressions can be justified by
more systematic studies~(see e.g. ref.~\cite{lprb}):
$f^\L(p)$ possesses an asymptotic expansion
involving only powers of the form $p^n$ and $p^{D/2-1+n}$, for $n=0,1,\dots$
Whenever $D=2,4,\dots$ is an even integer,
the two sequences of exponents merge, giving rise to logarithmic corrections,
which shall be discarded throughout the following.

In low enough dimension ($D<2$), $f^\L(p)$ diverges as $p\to0$.
As a consequence, for any finite temperature,
$g^\L(T,p)$ has a pole at some positive value of~$p$, denoted by $1/\taueq$,
away from the cut of $f^\L(p)$.
Hence
\beq
g(T,t)\comport{\sim}{t\to\infty}\e^{t/\taueq},
\label{ghigh}
\eeq
and therefore, as further analyzed below,
the system relaxes exponentially fast to equilibrium,
with a finite relaxation time $\taueq$.
The latter diverges as the zero-temperature phase transition is approached, as
\beq
\taueq\comport{\approx}{T\to0}
\left(2(8\pi)^{-D/2}\Gamma(1-D/2)T\right)^{-2/(2-D)}.
\eeq

In high enough dimension ($D>2$), $f^\L(p=0)=A_1$ is finite,
so that the pole of $g^\L(T,p)$ hits the cut of $f^\L(p)$
at $p=0$ at a finite critical temperature
\beq
T_c=\frac{1}{2A_1}=\left(\int\dq\,\frac{1}{\omega(\q)}\right)^{-1}.
\label{tc}
\eeq
As $T\to T_c^+$, the relaxation time $\taueq$ diverges according to
\beqa
2<D<4:&&\taueq\comport{\approx}{T\to
T_c^+}\left(\frac{2(8\pi)^{-D/2}\abs{\Gamma(1-D/2)}T_c^2}
{T-T_c}\right)^{2/(D-2)},\cr\cr
D>4:&&\taueq\comport{\approx}{T\to T_c^+}\frac{2A_2T_c^2}{T-T_c}.
\label{taueqas}
\eeqa
\setcounter{footnote}{0}
Note that these equations can be recast into the form
$\taueq\sim(T-T_c)^{-\nu z_c}$,
where $\nu$ is the critical exponent of the correlation length,
equal to $1/(D-2)$ for $2<D<4$ and to $1/2$ for $D>4$~\cite{baxter},
while $z_c$ is the dynamic critical exponent, equal to 2 in the present
case.\footnote{A summary of the values of static and dynamical exponents
appearing in this work is given in Table~1.}

We now discuss the asymptotic behavior of the function $g(T,t)$
according to temperature.
Throughout the following, we will assume that $D>2$,
so that the model has a ferromagnetic transition at a finite $T_c$,
given by eq.~(\ref{tc}).

\begin{itemize}

\item In the paramagnetic phase $(T>T_c)$,
$g(T,t)$ still grows exponentially, according to eq.~(\ref{ghigh}).

\item In the ferromagnetic phase $(T<T_c)$,
a careful analysis of eq.~(\ref{fglap}) yields
\beq
g(T,t)\comport{\approx}{t\to\infty}\frac{f(t)}{M_\eq^4}
\approx\frac{(8\pi t)^{-D/2}}{M_\eq^4},
\label{glow}
\eeq
where the spontaneous magnetization $M_\eq$ is given by~\cite{baxter}
\beq
M_\eq^2=1-\frac{T}{T_c}.
\eeq

\item At the critical point $(T=T_c)$, we obtain
\beqa
2<D<4:&&g(T_c,t)\comport{\approx}{t\to\infty}
(D-2)(8\pi)^{D/2-1}\sin[(D-2)\pi/2]\frac{t^{-(2-D/2)}}{T_c^2},\cr
D>4:&&g(T_c,t)\comport{\to}{t\to\infty}\frac{1}{4A_2T_c^2}.
\label{gcrit}
\eeqa

\end{itemize}

Finally, eqs.~(\ref{fglap}), (\ref{fldef}), (\ref{flist}),
and~(\ref{tc}) yield the following identities:
\beqa
\int_0^\infty f(t)\,\d t&=&\frac{1}{2T_c},\cr\cr
\int_0^\infty f(t)\,\e^{-t/\taueq}\,\d t&=&\frac{1}{2T}
{\hskip 1.8cm}(T>T_c),\cr\cr
\int_0^\infty g(T,t)\,\d t&=&\frac{1}{2T_cM_\eq^2}
{\hskip 1cm}(T<T_c).
\label{fgid}
\eeqa

We are now in a position to discuss the temporal behavior of
the equal-time correlation function in the different phases.
Its expression~(\ref{cq}) in Fourier space reads
\beq
C^\F(\q,t)=\frac{\e^{-2\omega(\q)t}}{g(T,t)}
\left(1+2T\int_0^t\e^{2\omega(\q)t_1}g(T,t_1)\,\d t_1\right),
\label{cqfg}
\eeq
using the definition~(\ref{gdef}) of $g(T,t)$.
We shall consider in particular the dynamical susceptibility
\beq
\chi(t)=\frac{1}{T}\sum_\x\mean{S_\0(t)S_\x(t)}=\frac{C^\F(\q=\0,t)}{T},
\eeq
for which eq.~(\ref{cqfg}) yields
\beq
\chi(t)=\frac{1}{g(T,t)}\left(\frac{1}{T}+2\int_0^t g(T,t_1)\,\d t_1\right).
\eeq
The asymptotic expressions~(\ref{ghigh}), (\ref{glow}),
and~(\ref{gcrit}) of $g(T,t)$ lead to the following predictions.

\begin{itemize}

\item In the paramagnetic phase $(T>T_c)$, the correlation function converges
exponentially fast to its equilibrium value, which has the Ornstein-Zernike
form
\beq
C^\F_\eq(\q)=\frac{T}{\omega(\q)+\xi_\eq^{-2}},
\label{cfeq}
\eeq
where the equilibrium correlation length $\xi_\eq$ is given by
\beq
\xi_\eq^2=2\taueq.
\eeq
The corresponding value of the equilibrium susceptibility is
$\chi_\eq=\xi_\eq^2=2\taueq$.
Eq.~(\ref{cfeq}) implies an exponential and isotropic fall-off
of correlations, of the form $C_{\x,\eq}\sim\e^{-|\x|/\xi_\eq}$,
at large distances and for $\xi_\eq$ large, i.e., $T$ close enough to $T_c$.

\item In the ferromagnetic phase $(T<T_c)$,
using the third of the identities~(\ref{fgid}),
we obtain a scaling form for the correlation function, namely
\beq
C^\F(\q,t)\approx M_\eq^2\,(8\pi t)^{D/2}\,\e^{-2\q^2t},
\eeq
or equivalently,
\beq
C_\x(t)\approx M_\eq^2\,\e^{-\x^2/(8t)},
\eeq
in the regime where $\x$ is large (i.e., $\q$ is small) and~$t$ is large.
Both the Gaussian profile of the correlation function,
and its scaling law involving one single diverging length scale
\beq
L(t)\sim t^{1/2},
\label{lt}
\eeq
reflect the diffusive nature of the coarsening process.
The growing length $L(t)$ can be interpreted as the characteristic
size of an ordered domain.
The dynamical susceptibility,
\beq
\chi(t)\approx\frac{M_\eq^2}{T}(8\pi t)^{D/2},
\label{chilow}
\eeq
grows as $\chi(t)\sim L(t)^D$,
or else as the volume explored by a diffusive process.

\item At the critical point $(T=T_c)$,
the equilibrium correlation function reads
\beq
C^\F_\eq(\q)\approx\frac{T_c}{\q^2},
\eeq
i.e.,
\beq
C_{\x,\eq}\approx\frac{\Gamma(D/2-1)}{4\,\pi^{D/2}}
\,\frac{T_c}{\abs{\x}^{D-2}}.
\eeq
These limiting expressions are reached
according to scaling laws of the form
\beqa
C^\F(\q,t)&\approx&C^\F_\eq(\q)\,\Phi(\q^2t),\cr
C_\x(t)&\approx&C_{\x,\eq}\,\Psi(\x^2/t),
\label{cxsca}
\eeqa
with
\beqa
2<D<4:&&\Phi(x)=2x\int_0^1\e^{-2x(1-z)}\,z^{D/2-2}\,\d z,\cr
&&\Psi(y)=\e^{-y/8},\cr\cr
D>4:&&\Phi(x)=1-\e^{-2x},\cr
&&\Psi(y)=\frac{1}{\Gamma(D/2-1)}\int_{y/8}^\infty
\e^{-z}\,z^{D/2-2}\,\d z.
\eeqa

The second expression of eq.~(\ref{cxsca}) has the general
scaling form for the equal-time correlation function
(see eq.~(\ref{cxtcrit})),
with the known value of the static exponent of correlations $\eta=0$
for the spherical model~\cite{baxter}, and with $z_c=2$, already found above.

The dynamical susceptibility grows linearly with time,
as $\chi(t)\approx\Phi'(0)\,t$, i.e.,
\beqa
2<D<4:&&\chi(t)\approx\frac{4}{D-2}\,t,\cr\cr
D>4:&&\chi(t)\approx2t.
\eeqa

\end{itemize}

\subsection{Two-time correlation function}

We now consider the two-time correlation function
\beq
C_{\x-\y}(t,s)=\mean{S_\x(t)S_\y(s)},
\eeq
with $0\le s$~(waiting time) $\le t$~(observation time).
Its Fourier transform $C^\F(\q,t,s)$ is defined as in eq.~(\ref{cqdef}).
Using eq.~(\ref{sq}), we obtain
\beq
C^\F(\q,t,s)=\frac{\e^{-\omega(\q)(t+s)}}{\sqrt{g(T,t)g(T,s)}}
\left(1+2T\int_0^{s}\e^{2\omega(\q)t_1}g(T,t_1)\,\d t_1\right),
\label{cqts}
\eeq
or else
\beq
C^\F(\q,t,s)=C^\F(\q,s)\,\e^{-\omega(\q)(t-s)}\sqrt\frac{g(T,s)}{g(T,t)},
\label{cqts2}
\eeq
using the expression~(\ref{cqfg}) for $C^\F(\q,s)$.

In the following, we shall be mostly interested in the two-time
autocorrelation function
\beq
C(t,s)\equiv C_\0(t,s)=\mean{S_\x(t)S_\x(s)}=\int\dq\,C^\F(\q,t,s),
\eeq
for which eq.~(\ref{cqts}) yields
\beq
C(t,s)=\frac{1}{\sqrt{g(T,t)g(T,s)}}
\left[f\!\left(\frac{t+s}{2}\right)
+2T\int_0^{s}f\!\left(\frac{t+s}{2}-t_1\right)g(T,t_1)\,\d t_1\right].
\eeq
The autocorrelation with the initial state assumes the simpler form
\beq
C(t,s=0)=\frac{f(t/2)}{\sqrt{g(T,t)}}.
\label{czero}
\eeq

The asymptotic expressions~(\ref{fdef}), (\ref{ghigh}), (\ref{glow}),
and~(\ref{gcrit}) of the functions $f(t)$ and $g(T,t)$
lead to the following predictions.

\begin{itemize}

\item In the paramagnetic phase $(T>T_c)$, as $s\to\infty$
with $\tau=t-s$ fixed, the system converges to its equilibrium state,
where the correlation function only depends on~$\tau$:
\beq
C(s+\tau,s)\comport{\to}{s\to\infty}
C_\eq(\tau)=T\int_\tau^\infty f(\tau_1/2)\,\e^{-\tau_1/(2\taueq)}\,\d\tau_1.
\label{ceq}
\eeq
This equilibrium correlation function decreases exponentially to zero
as $\e^{-\tau/(2\taueq)}$ when $\tau\to\infty$.
The initial value $C_\eq(\tau=0)=1$ is ensured by the second identity
of eq.~(\ref{fgid}).

\item In the ferromagnetic phase $(T<T_c)$, two regimes need to be considered.
In the first regime ($s\to\infty$ and $\tau$~fixed, i.e., $1\sim\tau\ll s$),
using again the identities~(\ref{fgid}), we obtain
\beq
C(s+\tau,s)\approx M_\eq^2+(1-M_\eq^2)C_\eqcr(\tau),
\label{cbeta}
\eeq
where we have set
\beq
C_\eqcr(\tau)=T_c\int_\tau^\infty f(\tau_1/2)\,\d\tau_1.
\label{cc0}
\eeq
This function, which corresponds to the $T\to T_c$ limit of eq.~(\ref{ceq}),
decreases only algebraically to zero when $\tau\to\infty$, as
\beq
C_\eqcr(\tau)\comport{\approx}{\tau\to\infty}
\frac{2(4\pi)^{-D/2}}{D-2}\,T_c\,\tau^{-(D/2-1)},
\label{cc1}
\eeq
as implied by eq.~(\ref{fdef}).
The first identity of eq.~(\ref{fgid}) ensures that
$C_\eqcr(\tau=0)=1$.

In the second regime,
where~$s$ and~$t$ are simultaneously large (i.e., $1\ll s\sim\tau$),
with arbitrary ratio
\beq
x=\frac{t}{s}=1+\frac{\tau}{s}\ge1,
\label{dxts}
\eeq
the correlation function obeys a scaling law of the form
\beq
C(t,s)\approx M_\eq^2\left(\frac{4ts}{(t+s)^2}\right)^{D/4}
\approx M_\eq^2\left(\frac{4x}{(x+1)^2}\right)^{D/4}.
\label{calpha}
\eeq
When $x\gg1$, this expression behaves as
\beq
C(t,s)\approx A\,M_\eq^2\,x^{-\lambda/2},
\eeq
which can be recast into
\beq
C(t,s)\sim M_\eq^2\,\left(\frac{L(t)}{L(s)}\right)^{-\lambda},
\label{fish}
\eeq
where $L(t)$ is the length scale defined in eq.~(\ref{lt}),
and $\lambda$ is the autocorrelation exponent (see section~3),
which is equal to $D/2$ in the present case,
in agreement with the result found in the $n\to\infty$ limit of the $O(n)$
model (see ref.~\cite{bray}, p.~386, and references therein).

Between these two regimes, the correlation function takes a plateau value
\beq
\qea=\lim_{\tau\to\infty}\lim_{s\to\infty}C(s+\tau,s)
=M_\eq^2=1-\frac{T}{T_c},
\label{qeadef}
\eeq
known as the Edwards-Anderson order parameter (see e.g. ref.~\cite{mezard}).

Hereafter we shall refer to the first regime ($1\sim\tau\ll s$)
as the {\it stationary} regime, and to the second one ($1\ll s\sim\tau$)
as the {\it scaling} (or {\it aging}) regime.
In the former, the system becomes stationary,
though without reaching thermal equilibrium,
because the system is coarsening.
In the latter regime, as said above, the system is aging.
It is possible to match these two kinds of behavior,
corresponding respectively to eq.~(\ref{cbeta}) and~(\ref{calpha}),
into a single expression:
\beq
C(t=s+\tau,s)\approx
(1-M_\eq^2)C_\eqcr(\tau)
+M_\eq^2\left(\frac{4ts}{(t+s)^2}\right)^{D/4},
\eeq
which is the sum of a term corresponding to the stationary contribution,
and a term corresponding to the aging one.
Let us finally recall that, in the context of glassy dynamics,
in a low-temperature phase, the first regime, where $C(t,s)>\qea$, is usually
referred to as the $\beta$ regime, while the second one,
where $C(t,s)<\qea$, is referred to as the $\alpha$ regime~\cite{revue}.

\item At the critical point $(T=T_c)$,
the same two regimes are to be considered.
However their physical interpretation is slightly different, since
the order parameter $M_\eq$ vanishes, and symmetry
between the phases is restored.

In the first regime $(1\sim\tau\ll s)$, the system again becomes stationary,
the autocorrelation function behaving as the $T\to T_c$ limit
of eq.~(\ref{ceq}), that is
\beq
C(s+\tau,s)\comport{\to}{s\to\infty}C_\eqcr(\tau),
\label{cc}
\eeq
which decreases algebraically to zero
when $\tau\to\infty$ (cf. eq.~(\ref{cc1})).
In the second regime ($1\ll s\sim\tau$),
the correlation function obeys a scaling law of the form
\beq
C(t,s)\approx T_c\,s^{-(D/2-1)}\,F(x),
\label{csca}
\eeq
where the scaling function $F(x)$ reads
\beqa
2<D<4:&&F(x)=\frac{4(4\pi)^{-D/2}}{(D-2)(x+1)}\,x^{1-D/4}(x-1)^{1-D/2},\cr\cr
D>4:&&F(x)=\frac{2(4\pi)^{-D/2}}{D-2}\left((x-1)^{1-D/2}-(x+1)^{1-D/2}\right).
\label{f}
\eeqa

In this regime the system is still aging, in the sense that $C(t,s)$
bears a dependence in both time variables.
However, the scaling of expression~(\ref{csca}) is different from that
found in the low-temperature phase (see eq.~(\ref{calpha})),
which depends on the ratio $x=t/s$ only.
The presence in eq.~(\ref{csca}) of an additional $s$-dependence
through the factor $s^{-(D/2-1)}$
can be interpreted as coming from the anomalous
dimension of the field $S_\x$ at $T_c$.
In the critical region one has indeed
$M_\eq\sim(T-T_c)^{\beta}\sim\xi_\eq^{-\beta/\nu}$.
Replacing $\xi_\eq$ by $s^{1/z_c}$ implies the replacement of
$M_\eq^2$ by $s^{-2\beta/\nu z_c}\sim s^{-(D-2+\eta)/z_c}$.
With $\eta=0$ and $z_c=2$, the factor $s^{-(D/2-1)}$ is thus recovered.
Note that the static hyperscaling relation $2\beta/\nu=D-2+\eta$
holds for $D<4$, while it is violated for $D>4$ (see Table~1).

Two limiting regimes are of interest.
First, for $x\to1$, i.e., $1\ll\tau\ll s$,
eq.~(\ref{csca}) matches eq.~(\ref{cc1}).
Second, for $x\gg1$, i.e., $1\ll s\ll t$, one gets
\beq
F(x)\approx B\,x^{-\lambda_c/z_c},
\label{fxxgd}
\eeq
where the autocorrelation exponent $\lambda_c$ (see section~3)
is equal to $3D/2-2$ if $2<D<4$,
and to~$D$ above four dimensions, in agreement with the result
found in ref.~\cite{janssen}.

We also quote for later reference the scaling law of the derivative
\beq
\frac{\dpar C(t,s)}{\dpar s}\approx T_c\,s^{-D/2}\,F_1(x),
\label{dcsca}
\eeq
with
\beq
F_1(x)=-\frac{D-2}{2}\,F(x)-x\,F'(x),
\eeq
i.e.,
\beqa
2<D<4:&&F_1(x)=(4\pi)^{-D/2}\frac{(D-2)(x+1)^2+2(x-1)^2}{(D-2)(x+1)^2}
\,x^{1-D/4}(x-1)^{-D/2},\cr\cr
D>4:&&F_1(x)=(4\pi)^{-D/2}\left((x-1)^{-D/2}+(x+1)^{-D/2}\right).
\label{f1}
\eeqa

\end{itemize}

\subsection{Two-time response function}

Suppose now that the system is subjected to a small magnetic field $H_\x(t)$,
depending on the site $\x$ and on time $t\ge0$ in an arbitrary fashion.
This amounts to adding to the ferromagnetic Hamiltonian~(\ref{ham})
a time-dependent perturbation of the form
\beq
\delta\H(t)=-\sum_\x H_\x(t)S_\x(t).
\eeq
The dynamics of the model is now given by the modified Langevin equation
\beq
\frac{\d S_\x}{\d t}=\sum_{\y(\x)}S_\y-\lambda(t)S_\x+H_\x(t)+\eta_\x(t).
\label{lanh}
\eeq

Causality and invariance under spatial translations imply that we have,
to first order in the magnetic field $H_\x(t)$,
\beq
\mean{S_\x(t)}=\int_0^t\d s\sum_\y R_{\x-\y}(t,s)H_\y(s)+\cdots
\label{sxt}
\eeq
This formula defines the two-time response function $R_{\x-\y}(t,s)$
of the model.
A more formal definition reads
\beq
R_{\x-\y}(t,s)=\left.\frac{\delta\mean{S_\x(t)}}
{\delta H_\y(s)}\right\vert_{\{H_\x(t)=0\}}.
\eeq

The solution to eq.~(\ref{lanh}) reads, in Fourier space,
\beq
S^\F(\q,t)=\e^{-\omega(\q)t-\z(t)}
\left(S^\F(\q,t=0)+\int_0^t\e^{\omega(\q)t_1+\z(t_1)}
\left[H^\F(\q,t_1)+\eta^\F(\q,t_1)\right]\,\d t_1\right).
\eeq
It can be checked that the Lagrange function $\lambda(t)$,
and hence $z(t)$ and $\z(t)$,
remain unchanged, to first order in the magnetic field.
As a consequence, the two-time response function reads, in Fourier transform,
\beq
R^\F(\q,t,s)=\left.\frac{\delta\mean{S^\F(\q,t)}}{\delta H^\F(\q,s)}
\right\vert_{\{H_\x(t)=0\}}
=\e^{-\omega(\q)(t-s)}\sqrt\frac{g(T,s)}{g(T,t)}
\label{rqts}
\eeq
(cf. eq.~(\ref{cqts2})).
In the following, we shall be mostly interested in the diagonal component
of the response function, corresponding to coinciding points:
\beq
R(t,s)\equiv R_\0(t,s)
=\left.\frac{\delta\mean{S_\x(t)}}{\delta H_\x(s)}\right\vert_{\{H_\x(t)=0\}}
=\int\dq\,R^\F(\q,t,s).
\eeq
With the notations~(\ref{fdef}), (\ref{gdef}), eq.~(\ref{rqts}) yields
\beq
R(t,s)=f\!\left(\frac{t-s}{2}\right)\sqrt\frac{g(T,s)}{g(T,t)}.
\eeq
The response function at zero waiting time
assumes the simpler form~(cf.~eq.~(\ref{czero}))
\beq
R(t,s=0)=C(t,s=0)=\frac{f(t/2)}{\sqrt{g(T,t)}}.
\eeq

The asymptotic expressions~(\ref{fdef}), (\ref{ghigh}), (\ref{glow}),
and~(\ref{gcrit}) of $F(t)$ and $g(T,t)$ lead to the following predictions.

\begin{itemize}

\item In the paramagnetic phase $(T>T_c)$,
at equilibrium, the response function only depends on $\tau$, according to
\beq
R_\eq(\tau)=f(\tau/2)\,\e^{-\tau/(2\taueq)}.
\label{req}
\eeq
Moreover, it is related to the equilibrium correlation function
$C_\eq(\tau)$ of eq.~(\ref{ceq})
by the fluctuation-dissipation theorem~(\ref{fdt}), as it should.

\item In the ferromagnetic phase $(T<T_c)$,
the two regimes defined in the previous section for the case of
the autocorrelation function are still to be considered.
In the stationary regime ($1\sim\tau\ll s$),
the response function behaves as the $T\to T_c$ limit of eq.~(\ref{req}),
namely
\beq
R_\eqcr(\tau)=f(\tau/2)=-\frac{1}{T_c}\frac{\d C_\eqcr(\tau)}{\d\tau},
\label{rc}
\eeq
so that the \fd theorem is valid.

On the contrary, in the scaling regime ($1\ll s\sim t$),
the response function has the form
\beq
R(t,s)\approx(4\pi(t-s))^{-D/2}(t/s)^{D/4}=
(4\pi s)^{-D/2}\,(x-1)^{-D/2}x^{D/4},
\label{ralpha}
\eeq
which, when compared to
the corresponding expression~(\ref{calpha}) for the autocorrelation function,
demonstrates the violation of the \fd theorem (see section~2.5).

\item At the critical point $(T=T_c)$,
in the stationary regime $(1\sim\tau\ll s)$,
the response function still behaves as in eq.~(\ref{rc}),
so that the \fd theorem still holds.
In the scaling regime ($1\ll s\sim t$),
the response function obeys a scaling law of the form
\beq
R(t,s)\approx s^{-D/2}\,F_2(x),
\label{rsca}
\eeq
where the scaling function $F_2(x)$ reads
\beqa
2<D<4:&&F_2(x)=(4\pi)^{-D/2}x^{1-D/4}(x-1)^{-D/2},\cr
D>4:&&F_2(x)=(4\pi)^{-D/2}(x-1)^{-D/2}.
\label{f2}
\eeqa

Again two limiting regimes are of interest.
For $x\to1$, the scaling result~(\ref{f2}) matches eq.~(\ref{rc}).
For $x\gg 1$, one finds the same power-law fall-off
for the functions $F(x)$, $F_1(x)$, and $F_2(x)$, that is
\beq
F(x)\sim F_1(x)\sim F_2(x)\sim x^{-\lambda_c/z_c}
\eeq
(see sections~2.5 and~3).

\end{itemize}

\subsection{Fluctuation-dissipation ratio}

As already mentioned in the introduction,
the violation of the \fd theorem~(\ref{fdt}) out of thermal equilibrium
can be characterized by the \fd ratio $X(t,s)$, defined in eq.~(\ref{dx}).
In the case of the spherical model,
the results derived so far yield at once the following predictions.

\begin{itemize}

\item In the paramagnetic phase $(T>T_c)$,
the system converges to an equilibrium state, where the \fd theorem holds.
In other words, the \fd ratio converges toward its equilibrium value
\beq
X_\eq=1.
\eeq

\item In the ferromagnetic phase $(T<T_c)$,
the \fd theorem~(\ref{fdt}) is only valid in the stationary regime
($1\sim\tau\ll s$).
On the contrary, in the scaling regime ($1\ll s\sim\tau$),
the results~(\ref{calpha}) and~(\ref{ralpha})
imply that the \fd ratio falls off as
\beq
X(t,s)\approx\frac{(8\pi)^{-D/2}}{D}\,\frac{4T}{M_\eq^2}\,
\left(\frac{x+1}{x-1}\right)^{D/2+1}s^{-(D/2-1)}.
\label{vanis}
\eeq
In particular, the limit \fd ratio introduced in eq.~(\ref{xinfdef}) reads
\beq
X_\infty=0.
\eeq

\item At the critical point $(T=T_c)$,
the scaling laws~(\ref{dcsca}) and~(\ref{rsca})
imply that the \fd ratio $X(t,s)$ becomes asymptotically
a smooth function of the time ratio $x=t/s$:
\beq
X(t,s)\comport{\approx}{t,s\to\infty}\X(x)=\frac{F_2(x)}{F_1(x)},
\label{xsca}
\eeq
i.e., explicitly,
\beqa
2<D<4:&&\X(x)=\frac{1}{1+\frad{2}{D-2}\left(\frad{x-1}{x+1}\right)^2},\cr\cr
D>4:&&\X(x)=\frac{1}{1+\left(\frad{x-1}{x+1}\right)^{D/2}}.
\label{x}
\eeqa
The scaling law~(\ref{xsca}) interpolates between the equilibrium behavior
\beq
\X(x)\comport{\to}{x\to1}X_\eq=1
\eeq
in the stationary regime of relatively short time differences,
and a non-trivial limit value
\beq
\X(x)\comport{\to}{x\to\infty}X_\infty
\eeq
at large time differences, given by
\beqa
2<D<4:&&X_\infty=\frac{D-2}{D},\cr\cr
D>4:&&X_\infty=\frac{1}{2}.
\label{xexpl}
\eeqa

\end{itemize}

Further comments on the scaling behavior of the \fd ratio
will be made in section~3.2.

\def\figun{
\vskip 8.5cm{\hskip .8cm}
\includegraphics{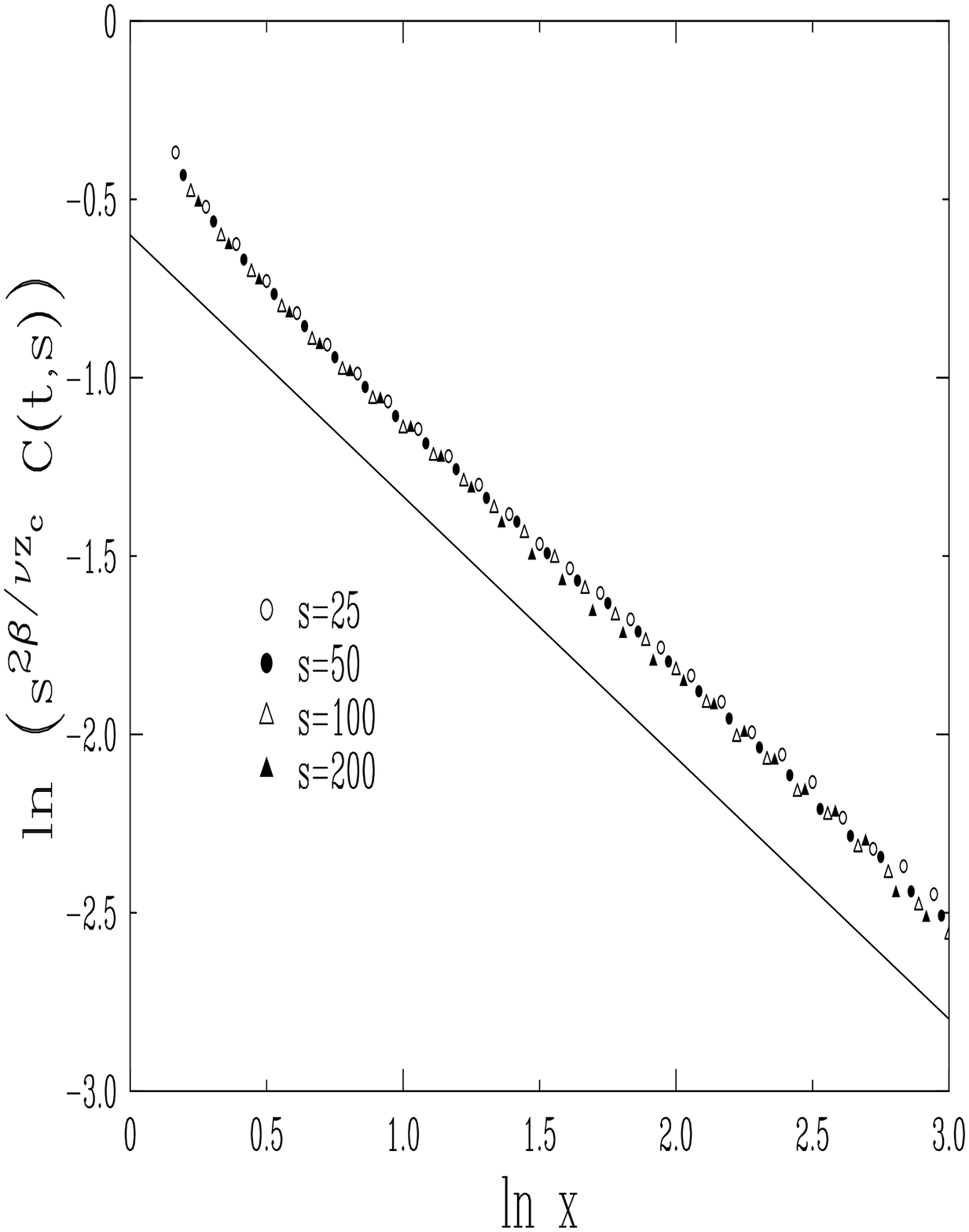}

\noindent {\bf Figure~1:}
Log-log plot of the critical autocorrelation function $C(t,s)$
of the two-di\-men\-si\-o\-nal Ising model,
against time ratio $x=t/s$, for several values of the waiting time~$s$.
Data are multiplied by $s^{2\beta/\nu z_c}$,
in order to demonstrate collapse into the scaling function $F(x)$
of eq.~(\ref{gc}).
Straight line: exponent $-\lambda_c/z_c\approx-0.73$
of the fall-off at large~$x$.
}

\def\figde{
\vskip 8.5cm{\hskip .8cm}
\includegraphics{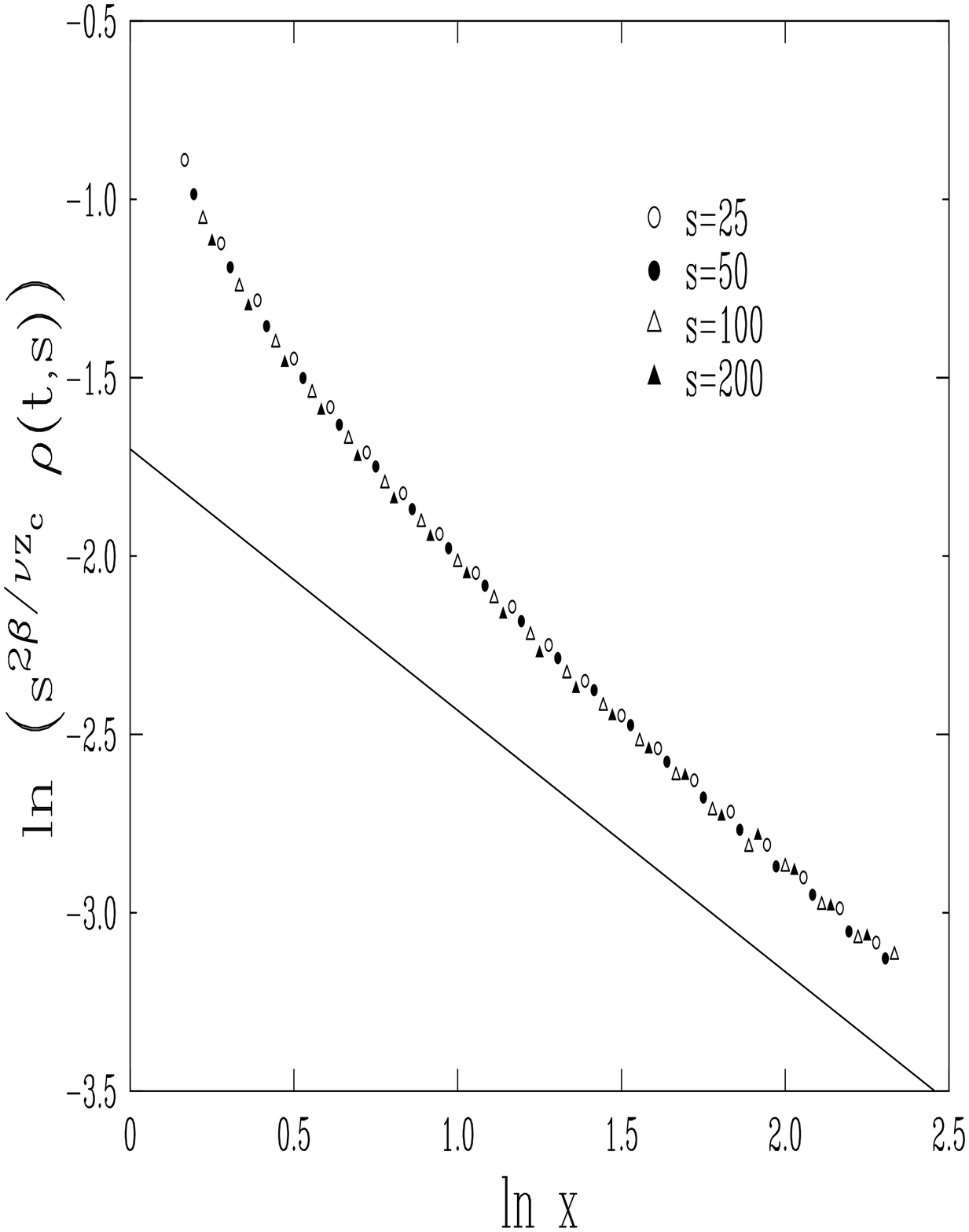}

\noindent {\bf Figure~2:}
Log-log plot of the critical integrated response function $\rho(t,s)$
of the two-dimensional Ising model,
against time ratio $x=t/s$, for several values of the waiting time~$s$.
Data are multiplied by $s^{2\beta/\nu z_c}$,
in order to demonstrate collapse into the scaling function $F_3(x)$
of eq.~(\ref{grint}).
Straight line: exponent $-\lambda_c/z_c\approx-0.73$
of the fall-off at large~$x$.
}

\def\figtr{
\vskip 8.5cm{\hskip .8cm}
\includegraphics{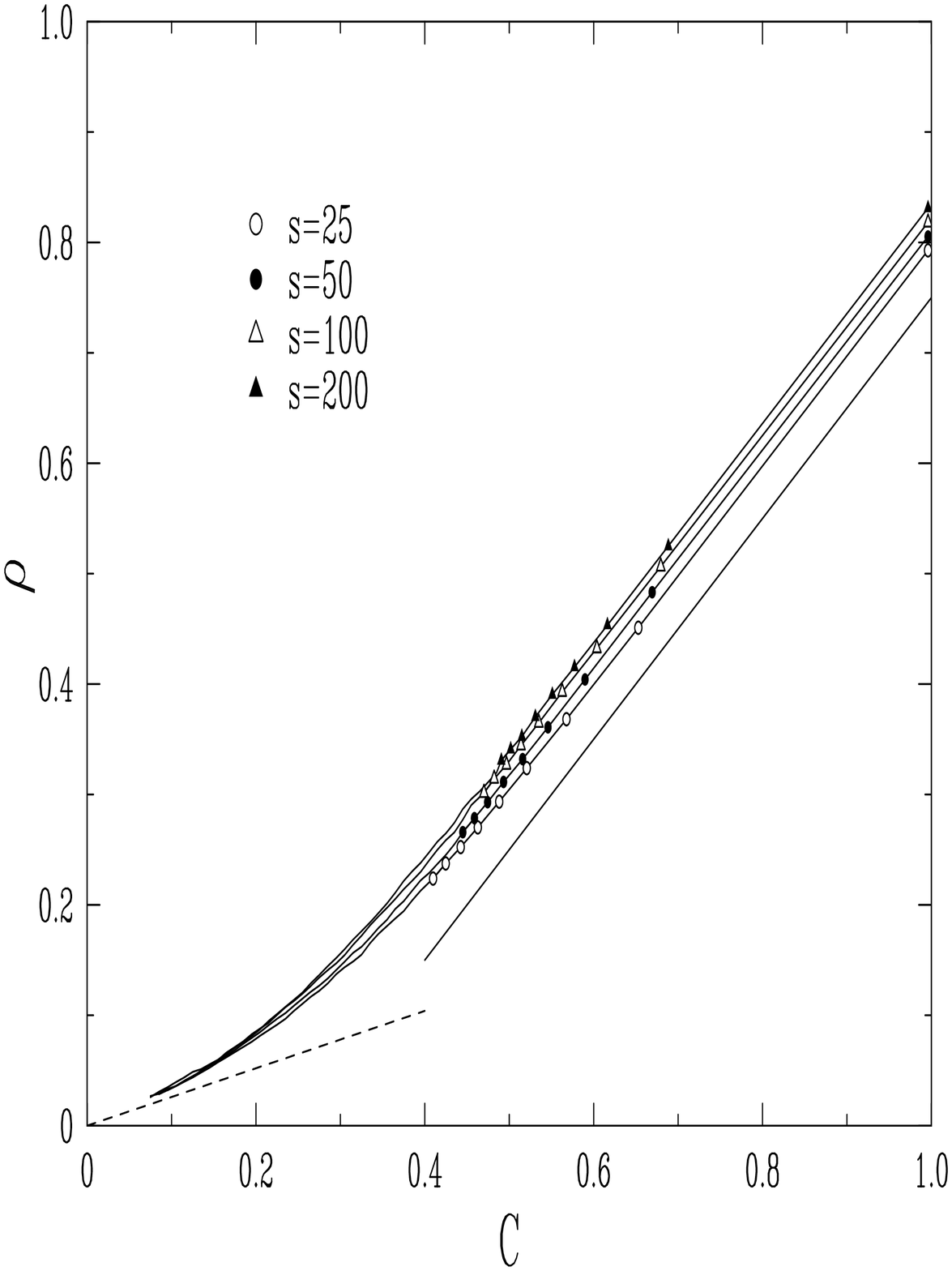}

\noindent {\bf Figure~3:}
Parametric plot of the integrated response $\rho(t,s)$
against the autocorrelation $C(t,s)$, using the data of Figures~1 and~2.
Symbols: data for integer time differences $\tau=t-s=0,\dots,8$.
Full line: unit slope corresponding to the \fd theorem
in the stationary regime.
Dashed line: limit slope $X_\infty=0.26$ (see text and Figures~4 and~5).
}

\def\figqu{
\vskip 8.5cm{\hskip .8cm}
\includegraphics{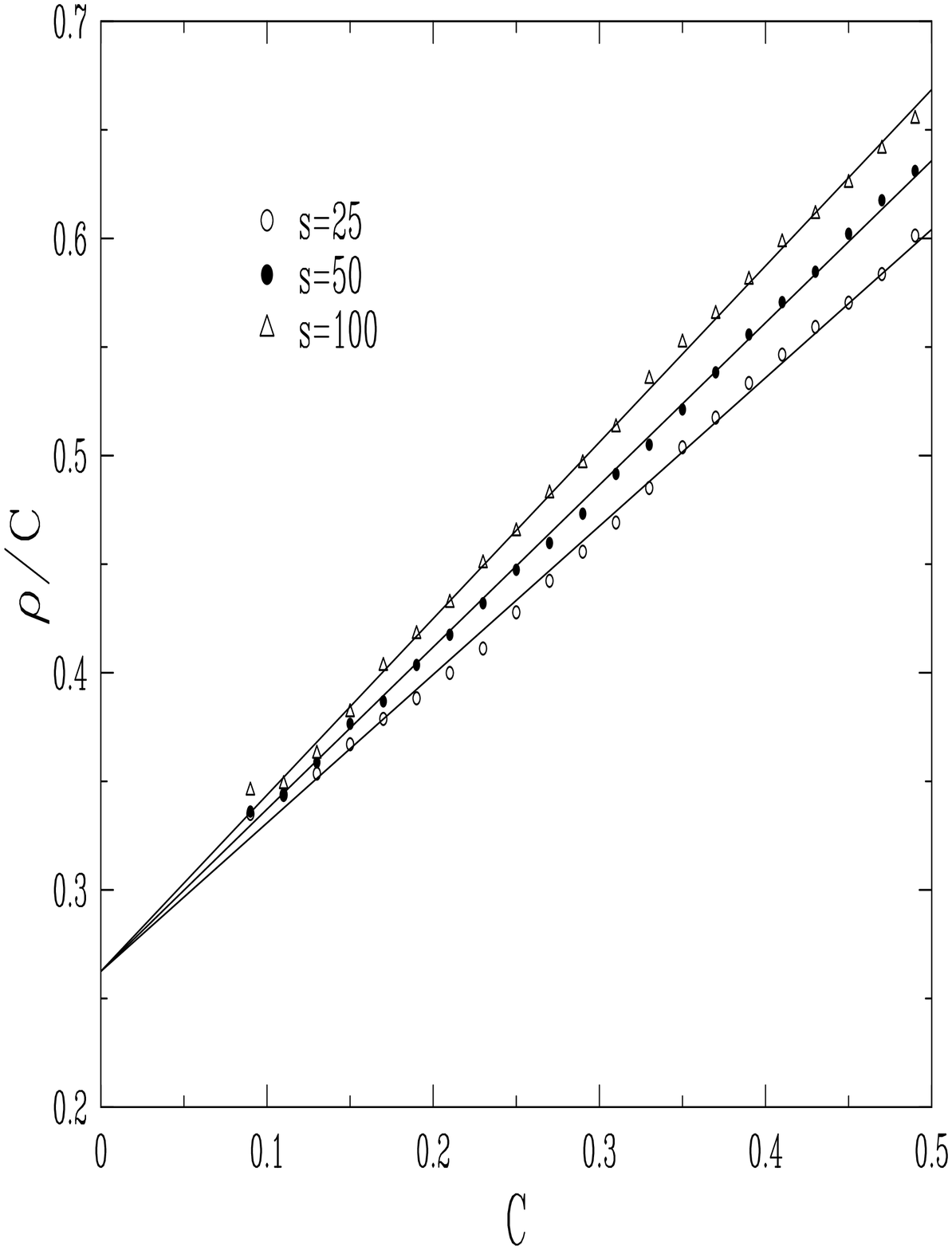}

\noindent {\bf Figure~4:}
Parametric plot of the ratio $\rho/C$ against~$C$.
Straight lines: constrained least-square fit with common intercept,
yielding $X_\infty\approx0.262$.
}

\def\figci{
\vskip 8.5cm{\hskip .8cm}
\includegraphics{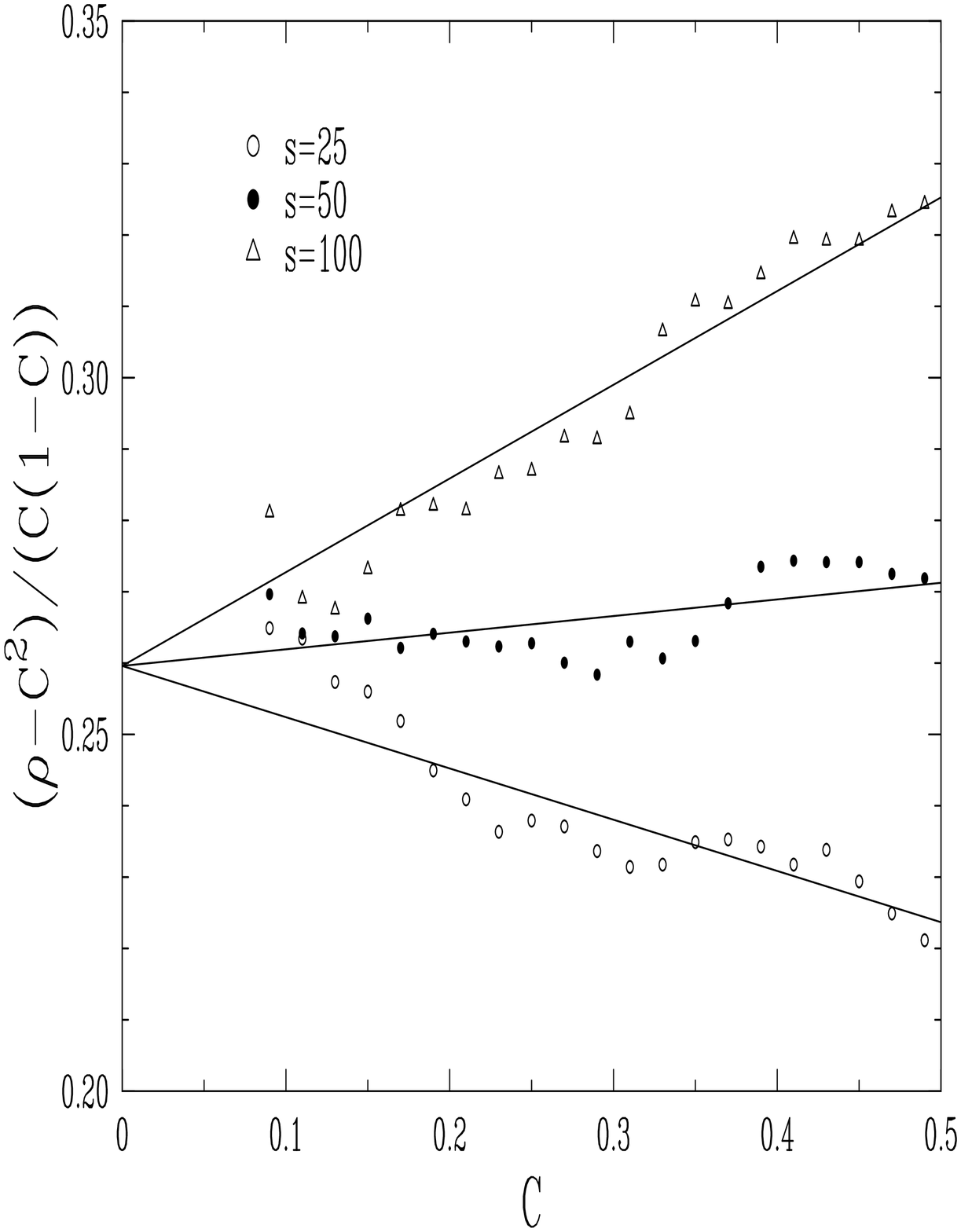}

\noindent {\bf Figure~5:}
Parametric plot of the combination $(\rho-C^2)/(C(1-C))$ against~$C$.
Straight lines: constrained least-square fit yielding $X_\infty\approx0.260$.
}

\section{The generic situation}

\subsection{Aging below $T_c$}

Let us first briefly sketch the description
of the dynamical behavior of a ferromagnetic system
quenched from a disordered initial state
to a temperature $T<T_c$~\cite{langer,bray,cd,barrat,berthier}.

In the scaling regime ($1\ll s\sim t$), the autocorrelation $C(t,s)$
is expected to be a function of the ratio $L(t)/L(s)$ only,
where the length scale $L(t)\sim t^{1/z}$ is the characteristic size
of an ordered domain, and~$z$ is the growth exponent,
equal to 2 for non-conserved dynamics.
More precisely,
\beq
C(t,s)=M_\eq^2\,f(t/s),
\label{lo1}
\eeq
where the scaling function~$f$ is temperature independent.
Furthermore we have, for $x=t/s\gg1$, i.e., $1\ll s\ll t$,
\beq
f(x)\approx A\,x^{-\lambda/z},
\label{lo2}
\eeq
where $\lambda$ is the autocorrelation exponent~\cite{fisher}.
For the spherical model, eqs.~(\ref{calpha}) and~(\ref{fish})
match eqs.~(\ref{lo1}) and~(\ref{lo2}), with $\lambda=D/2$ and $z=2$.

As a consequence, we have
\beq
\frac{\dpar C(t,s)}{\dpar s}\approx\frac{M_\eq^2}{s}\,f_1(x),
\label{lo3}
\eeq
with $f_1(x)=-xf'(x)$, so that, when $x\gg1$,
\beq
f_1(x)\approx A_1\,x^{-\lambda/z},
\label{lo5}
\eeq
with $A_1=A\,\lambda/z$.

Although the situation of the response $R(t,s)$ is less clear-cut,
it is however reasonable to make the scaling assumption
\beq
R(t,s)\approx s^{-1-a}\,f_2(x),
\label{gr2}
\eeq
where $a>0$ is an unknown exponent, and again with the behavior
\beq
f_2(x)\approx A_2\,x^{-\lambda/z}
\eeq
when $x\gg1$.

The scaling law~(\ref{gr2}) holds for the spherical model,
with $a=D/2-1$, as can be seen from eq.~(\ref{ralpha}).
Furthermore, for non-conserved dynamics,
at least in the case of a discrete broken symmetry,
like e.g. in the Ising model,
it has been argued~\cite{berthier,bray2} that the integrated response
$\rho(t,s)$ (to be defined in eq.~(\ref{rhodef})),
scales as $\rho(t,s)\sim L(s)^{-1}\,\varphi(L(t)/L(s))$.
This corresponds to eq.~(\ref{gr2}) with $a=1/z=1/2$.

The scaling laws~(\ref{lo3}), (\ref{gr2}) imply
\beq
X(t,s)\approx s^{-a}\,g(x),
\eeq
with $g(x)=(T/M_\eq^2)\,f_2(x)/f_1(x)$,
in agreement with eq.~(\ref{vanis}) for the spherical model, and especially
\beq
X_\infty=0.
\eeq

\subsection{Aging at $T_c$}

Let us now turn to the situation where a ferromagnetic
system is quenched from a disordered initial state to its critical point.

In such a circumstance, spatial correlations develop in the system,
just as in the critical state,
but only over a length scale which grows like $t^{1/z_c}$,
where $z_c$ is the dynamic critical exponent.
For example the equal-time correlation function has the scaling form
\beq
C_\x(t)=|\x|^{-2\beta/\nu}\,\phi\!\left(|\x|/t^{1/z_c}\right),
\label{cxtcrit}
\eeq
where $\beta$ and $\nu$ are the usual static critical exponents.
The scaling function $\phi(x)$ goes to a constant for $x\to0$,
while it falls off exponentially to zero for $x\to\infty$,
i.e., on scales smaller than $t^{1/z_c}$ the system looks critical,
while on larger scales it is disordered.
This behavior is illustrated in the case of the spherical model
by eq.~(\ref{cxsca}),
corresponding to $2\beta/\nu=D-2$ and $z_c=2$ in eq.~(\ref{cxtcrit}).

In the scaling region of the two-time plane,
where both times~$s$ and~$t$ are large and comparable ($1\ll s\sim t$),
with arbitrary ratio $x=t/s$,
the two-time autocorrelation function $C(t,s)$
is expected to obey a scaling law of the form
(see the discussion below eq.~(\ref{f}), and ref.~\cite{janssen})
\beq
C(t,s)\approx s^{-2\beta/\nu z_c}\,F(x).
\label{gc}
\eeq
When both time scales are well separated $(1\ll s\ll t$, i.e., $x\gg1$),
the scaling function $F(x)$ falls off as
\beq
F(x)\approx B\,x^{-\lambda_c/z_c},
\label{gfas}
\eeq
where $\lambda_c$ is the critical autocorrelation exponent~\cite{huse},
related to the (magnetization) initial-slip critical exponent
$\Theta_c$~\cite{janssen} by $\lambda_c=D-z_c\Theta_c$.

We thus have
\beq
\frac{\dpar C(t,s)}{\dpar s}\approx s^{-1-2\beta/\nu z_c}\,F_1(x),
\label{gdc}
\eeq
with
$F_1(x)=-(2\beta/\nu z_c)F(x)-xF'(x)$, so that, when $x\gg1$,
\beq
F_1(x)\approx B_1\,x^{-\lambda_c/z_c},
\label{gf1as}
\eeq
with
\beq
B_1=\frac{\nu\lambda_c-2\beta}{\nu z_c}\,B.
\label{bb1}
\eeq

For the spherical model, eqs.~(\ref{csca}), (\ref{fxxgd}), and~(\ref{dcsca})
respectively match eqs.~(\ref{gc}), (\ref{gfas}), and~(\ref{gdc}), with
$\lambda_c=3D/2-2$ if $D<4$, and $\lambda_c=D$ if $D>4$ (see Table~1).

The similarity between the results~(\ref{dcsca}) and~(\ref{rsca}),
obtained in the case of the spherical model,
strongly suggests that $\dpar C(t,s)/\dpar s$
and $R(t,s)$ behave similarly in the generic case,
i.e., one is lead to hypothesize
that a scaling law of the form~(\ref{gdc}),
with the same power-law fall-off~(\ref{gf1as}),
holds for the response, that is
\beq
R(t,s)\approx\frac{1}{T_c}\,s^{-1-2\beta/\nu z_c}\,F_2(x),
\label{gr}
\eeq
with, when $x\gg1$,
\beq
F_2(x)\approx B_2\,x^{-\lambda_c/z_c}.
\label{gf2as}
\eeq

The scaling laws~(\ref{gdc}) and~(\ref{gr}) imply then that
the \fd ratio only depends on the
time ratio~$x$ throughout the scaling region:
\beq
X(t,s)\approx\X(x)=\frac{F_2(x)}{F_1(x)},
\label{xgen}
\eeq
where the scaling function $\X(x)$ is universal.
It appears indeed as a dimensionless combination of scaling functions.
In turn, the hypothesis~(\ref{gf2as}) implies that the limit \fd ratio reads
\beq
X_\infty=\X(\infty)=\frac{B_2}{B_1}.
\label{xr}
\eeq
This number thus appears as a dimensionless amplitude ratio,
in the usual sense of critical phenomena.
It is therefore a novel universal quantity
of non-equilibrium critical dynamics, as already claimed in ref.~\cite{gl}.

In the case of the spherical model,
the analytical treatment of section~2 corroborates the above analysis,
and yields the quantitative predictions~(\ref{x}) and~(\ref{xexpl}).

In order to perform a numerical evaluation of $X_\infty$,
one needs to measure the response.
A convenient way to do so is to
measure instead the dimensionless integrated response function
\beq
\rho(t,s)=T\int_0^s R(t,u)\,\d u.
\label{rhodef}
\eeq
By eq.~(\ref{sxt}), this quantity is proportional to the thermoremanent
magnetization $M_\TRM$,
i.e., the magnetization of the system at time~$t$
obtained after applying a small magnetic field~$h$, uniform and constant,
between $t=0$ and $t=s$:
\beq
M_\TRM(t,s)\approx\frac{h}{T}\,\rho(t,s).
\eeq
The thermoremanent magnetization is a natural quantity
to measure experimentally in spin glasses~\cite{revue},
and it is also accessible to numerical simulations,
for systems with and without quenched randomness (see section~3.3).

The scaling law~(\ref{gr}) for the response function implies
\beq
\rho(t,s)\approx s^{-2\beta/\nu z_c}\,F_3(x),
\label{grint}
\eeq
with $F_2(x)=-(2\beta/\nu z_c)F_3(x)-xF_3'(x)$, so that, when $x\gg1$,
\beq
F_3(x)\approx B_3\,x^{-\lambda_c/z_c},
\label{gf3as}
\eeq
with
\beq
B_3=\frac{\nu z_c}{\nu\lambda_c-2\beta}\,B_2.
\label{b2b3}
\eeq

A clear representation of the evolution of $X(t,s)$ in time is provided by
the parametric plot of $\rho(t,s)$ against $C(t,s)$,
obtained by varying~$t$ at fixed~$s$~\cite{x1,barrat,berthier}.
For well-separated times in the scaling regime (i.e., $1\ll s\ll t$),
the common power-law behavior ~(\ref{gfas}), (\ref{gf1as}), (\ref{gf2as}),
and~(\ref{gf3as}) implies that the limit \fd ratio
has the alternative expression
\beq
X_\infty=\frac{B_3}{B},
\label{xrho}
\eeq
which is equivalent to eq.~(\ref{xr}),
due to eqs.~(\ref{bb1}) and~(\ref{b2b3}).
In other words, the relationship~(\ref{dx}) also holds in integral form,
that is
\beq
\rho(t,s)\approx X_\infty\,C(t,s),
\label{rhosca}
\eeq
in the regime $1\ll s\ll t$.
The limit \fd ratio can thus be measured as the slope of the parametric plot
in the scaling region, i.e., near the origin of the $C-\rho$ plane.
Eq.~(\ref{rhosca}) is expected to hold as long
as~$C$ and $\rho$ are much smaller than the crossover scale
\beq
C^*(s)=C(2s,s)\sim s^{-2\beta/\nu z_c},
\label{sizecrit}
\eeq
corresponding to $\tau=s$.
This quantity provides a measure of the size of the critical region,
giving thus a quantitative definition of the critical analogue of $M_\eq^2$,
involved in the discussion below eq.~(\ref{f}).

\subsection{The two-dimensional Ising model: numerical simulations}

In order to check the validity of the scaling analysis
made in the previous section, beyond the case of the spherical model,
we have performed numerical simulations on the ferromagnetic Ising model
on the square lattice,
evolving under heat-bath (Glauber) dynamics at its critical temperature
$T_c=2/\ln(1+\sqrt{2})\approx2.2692$, starting from a disordered initial state.
The rules of the dynamics are as follows.
Consider a finite system, consisting of $N=L^2$ spins $\sigma_\x=\pm1$
situated at the vertices $\x$ of a square lattice,
with periodic boundary conditions.
The Ising Hamiltonian reads
\beq
\H=-\sum_{(\x,\y)}\sigma_\x\sigma_\y,
\eeq
where the sum runs over pairs of neighboring sites.
Heat-bath dynamics consists in
updating the spins $\sigma_\x(t)$ according to the stochastic rule
\beq
\sigma_\x(t)\to\left\{\matrix{
+1&\mbox{with prob.}\,\frad{1+\tanh(h_\x(t)/T_c)}{2},\hfill\cr\cr
-1&\mbox{with prob.}\,\frad{1-\tanh(h_\x(t)/T_c)}{2},\hfill\cr}\right.%}
\eeq
where the local field $h_\x(t)$ acting on $\sigma_\x(t)$ reads
\beq
h_\x(t)=\sum_{\y(\x)}\sigma_\y(t),
\label{hlocdef}
\eeq
with $\y(\x)$ denoting the four neighbors of site $\x$.

Let us give a brief summary of known facts on the dynamics of the Ising model.
For $T<T_c$, numerical studies have shown that the scaling
forms~(\ref{lo1}) and~(\ref{lo2}) hold,
with $z=2$ (non-conserved dynamics) and $\lambda\approx1.25$~\cite{fisher}.
The integrated response function (in another form,
known as the ZFC magnetization) has been measured in ref.~\cite{barrat}.
At $T=T_c$, the dynamic critical exponent reads $z_c\approx2.17$~\cite{blote},
and the autocorrelation exponent $\lambda_c\approx1.59$~\cite{huse,grass}.

Our aim is now to verify the hypotheses made in section~3.2,
especially the scaling laws~(\ref{gr}) and~(\ref{grint})
for the response function,
and to demonstrate the existence of a non-trivial limit $X_\infty$.

Computing $C(t,s)$ with good statistics is rather easy,
while the computation of $\rho(t,s)$ requires more effort.
We have followed the lines of the method introduced in ref.~\cite{barrat}.
In order to isolate the diagonal component of the response function,
a quenched, spatially random magnetic field,
is applied to the system from $t=0$ to $t=s$.
This magnetic field is of the form $H_\x=h_0\eps_\x$,
with a constant small amplitude $h_0$, and a quenched random modulation,
$\eps_\x=\pm1$ with equal probability, independently at each site $\x$.
The heat-bath dynamical rules are modified by adding up
the magnetic field $H_\x$ to the local field $h_\x(t)$ of eq.~(\ref{hlocdef}).
We have then
\beq
\overline{\langle\eps_\x\sigma_\x(t)\rangle}
=h_0\int_0^s R(t,u)\,\d u=\frac{h_0}{T}\,\rho(t,s)=M_{\rm TRM}(t,s),
\eeq
where the bar means an average with respect to
the distribution of the modulation $\eps_\x$ of the magnetic field.

\figun

We have first checked the validity of the scaling laws~(\ref{gc}),
and especially~(\ref{grint}).
Figures~1 and~2 respectively show log-log plots of the autocorrelation function
$C(t,s)$ and of the corresponding integrated response function $\rho(t,s)$,
against the time ratio $x=t/s$, for several values of the waiting time~$s$.
For each value of~$s$, the simulations are run up to $t/s=10$,
and data are averaged over at least 500 independent samples
of size $300\times300$.
For the response function, the amplitude of the quenched magnetic field
reads $h_0=0.05$.
Multiplying the data by $s^{2\beta/\nu z_c}$,
with $2\beta/\nu z_c\approx0.115$, gives good data collapse,
thus producing a plot of the scaling functions $F(x)$ and $F_3(x)$.
The data follow a power-law fall-off at large values of~$x$,
with a slope in good agreement with the value
$-\lambda_c/z_c\approx-0.73$, shown on the plots as a straight line.

We then turned to an investigation of the parametric plot
of these data in the $C-\rho$ plane.
At the qualitative level, this plot, shown in Figure~3
for several values of the waiting time~$s$, confirms our expectations.
The stationary regime $(1\sim\tau\ll s$, i.e., roughly speaking, $C>C^*(s))$,
corresponds to the right part of the plot.
The symbols show the data for small integer values of the time difference,
$\tau=t-s=0,\dots,8$, illustrating the fast decay of correlation
and integrated response in the stationary regime.
The rightmost points, corresponding to $\tau=0$, i.e., $C=C(s,s)=1$,
are compatible with the scaling law
$1-\rho(s,s)\sim C^*(s)\sim s^{-2\beta/\nu z_c}$.
The validity of the \fd theorem is testified
by the unit slope of this part of the plot, shown as a full straight line.
The aging regime $(1\ll s\sim t$, i.e., roughly speaking, $C<C^*(s))$,
corresponds to the left part of the plot.
As expected, the data crossover toward a non-trivial slope,
equal to the limit \fd ratio $X_\infty$.
The dashed line shows the slope $X_\infty=0.26$,
obtained by the analysis described below.

\figde

In order to obtain a quantitative prediction of the limit \fd ratio $X_\infty$,
we have followed two approaches.
Figure~4 depicts the local slope of the plot of Figure~3,
i.e., the ratio $\rho/C$, against~$C$, in the aging regime.
The data for the largest available waiting time $s=200$ have been discarded
from the analysis because they appear as too noisy on that scale.
The data look pretty linear all over the range presented in the plot.
This precocious scaling is due to the fact that the exponent
$2\beta/\nu z_c\approx0.115$ is small.
Hence the size of the critical region,
given by the estimate~(\ref{sizecrit}), is very large,
at least for waiting times~$s$ accessible to computer simulations.
We have indeed, for example, $C^*(100)=C(200,100)\approx0.24$.
The straight lines show a constrained least-square fit of the three
series of data, imposing a common intercept.
The value of this intercept yields the prediction $X_\infty\approx0.262$.

\figtr

We have also followed an alternative approach,
aiming at subtracting most of the deviations of the ratio $\rho/C$
with respect to its limit $X_\infty$ at $C\to0$.
This can be done by incorporating the known limit
of the stationary regime, i.e., $\rho\approx1$ as $C\to1$,
into a quadratic phenomenological formula:
$\rho\approx X_\infty C +(1-X_\infty) C^2$.
This formula can be rewritten as $X_\infty\approx(\rho-C^2)/(C(1-C))$,
suggesting to plot $(\rho-C^2)/(C(1-C))$ against~$C$,
instead of the mere ratio $\rho/C$.
This has been done in Figure~5.
As expected, the vertical scale has been considerably enlarged.
In return this procedure increases the statistical noise on the data points.
The straight lines again show a constrained least-square fit,
yielding $X_\infty\approx0.260$.

We can conclude from this numerical analysis that we have
\beq
X_\infty=0.26\pm0.01
\eeq
for the ferromagnetic Ising model in two dimensions.

\figqu

\section{Discussion}

In the present work we dealt with the dynamics of ferromagnetic
spin systems quenched from infinite temperature to their critical state.
This study, exemplified by the exact analysis of
the spherical model in any dimension $D>2$,
and by numerical simulations on the two-dimensional Ising model,
complements that of the Glauber-Ising chain,
presented in a companion paper~\cite{gl}.
The main results obtained in this work can be summarized as follows.

In such a non-equilibrium situation, these systems are aging
in the sense that their correlation and response functions
depend non-trivially on the waiting time~$s$
as well as on the observation time~$t$,
whenever these two times are simultaneously large.
The corresponding scaling laws (see eqs.~(\ref{gc}), (\ref{gdc}), (\ref{gr}),
and~(\ref{grint})), involve powers of~$s$,
related to the static anomalous dimension of the magnetization,
and universal scaling functions of the ratio $x=t/s$.
In the regime of large time separations, i.e., $1\ll s\ll t$ (or $x\gg1$),
these scaling functions fall off algebraically
with the common exponent $\lambda_c/z_c$.

The \fd ratio $X(t,s)$, characterizing the violation of the \fd theorem,
has a universal scaling form $\X(x)$,
and, for well-separated times in the aging regime,
it assumes a limit value $X_\infty$ equal to a dimensionless amplitude ratio
(see eqs.~(\ref{xr}) and~(\ref{xrho})).
Therefore, as announced in ref.~\cite{gl},
$X_\infty$ is a novel universal characteristic of critical dynamics,
which is intrinsically related to the non-equilibrium initial condition
of a critical quench from a disordered state.

\figci

The ferromagnetic models studied in the present work
turn out to have values of $X_\infty$ in the range
\beq
0\le X_\infty\le\frac{1}{2}.
\eeq
We have indeed $X_\infty=1-2/D$
if $2<D<4$, and $X_\infty=1/2$ for $D>4$, for the spherical model,
and $X_\infty\approx0.26\pm0.01$ for the two-dimensional Ising model.
Let us mention that preliminary simulations
on the three-dimensional Ising model yield $X_\infty\approx0.40$.
The backgammon, for which $X_\infty=1$~\cite{fr97,gl99},
thus belongs to another class of models.

The mean-field value
\beq
X_\infty^\mf=\frac{1}{2},
\eeq
obtained for the spherical model in dimension $D>4$,
also holds for a variety of models which are not mean-field-like,
including the Glauber-Ising chain~\cite{gl}
and the two-dimensional X-Y model at zero temperature~\cite{ckp}.

Let us finally discuss a few open questions.
It would be interesting to know whether there is an analogue for the present
case of the results found for models with discontinuous spin-glass transitions,
where the violation of the \fd theorem
is related to the configurational entropy~\cite{config}.
One would also like to know the status of the quantity $X_\infty$
for non-equilibrium systems with quenched disorder,
or for systems defined by dynamical rules without detailed balance.

In principle the limit \fd ratio $X_\infty$ could be calculated
by field-theoretical renormalization-group methods,
generalizing the computations done for universal amplitude ratios
in usual static critical phenomena~\cite{zinn},
as series in either $\eps=4-D$, or in $1/n$ for the $n$-component
Heisenberg model, the spherical model corresponding to the $n\to\infty$ limit.
The dimensionless time ratio $x=t/s$,
appearing in the two-time autocorrelation and response functions and \fd ratio,
is a temporal analogue of aspect ratios,
which play an important role in static critical phenomena
and finite-size scaling theory~\cite{fss}.
One may therefore wonder whether the latter,
and especially its latest developments involving conformal
and modular invariance,
could be used in order to put constraints on non-equilibrium critical dynamics.
Generalized symmetry groups, such as those introduced in ref.~\cite{henkel},
may also play a role in this issue.

\subsubsection*{Acknowledgements}

It is a pleasure for us to thank A.J. Bray and S.~Franz for interesting
discussions.

\newpage
\section*{Table and caption}

\vskip .6cm
\begin{center}
\begin{tabular}{|c|c|c|c|}
\hline
exponent&spherical $(2<D<4)$&spherical $(D>4)$&Ising $(D=2)$\\
\hline
$\eta$&$0$&$0$&$1/4$\\
$\beta$&$1/2$&$1/2$&$1/8$\\
$\nu$&$1/(D-2)$&$1/2$&$1$\\
\hline
$z$&$2$&$2$&$2$\\
$\lambda$&$D/2$&$D/2$&$\approx1.25$\\
\hline
$z_c$&$2$&$2$&$\approx2.17$\\
$\lambda_c$&$3D/2-2$&$D$&$\approx1.59$\\
$\Theta_c$&$1-D/4$&$0$&$\approx0.19$\\
\hline
\end{tabular}
\end{center}

\vskip .6cm
\noindent{\bf Table~1:}
Static and dynamical exponents of the ferromagnetic spherical model
and of the two-dimensional Ising model.
First group: usual static critical exponents $\eta$, $\beta$, and $\nu$
(equilibrium).
Second group: zero-temperature dynamical exponents~$z$ and $\lambda$
(coarsening below $T_c$).
Third group: dynamic critical exponents $z_c$, $\lambda_c$, and $\Theta_c$
(non-equilibrium critical dynamics).

\end{document}